\documentstyle[11pt,epsf]{article}
\makeatletter
\newcommand{\artsectnumbering}{%
\@addtoreset{equation}{section}
\renewcommand{\theequation}{\thesection.\arabic{equation}}}
\makeatother  
\newcommand{\al}{\alpha}

\newcommand{\bv}{\begin{verbatim}}
\newcommand{\de}{\delta}
\newcommand{\ch}{\chi}

\newcommand{\ep}{\epsilon}
\newcommand{\et}{\eta}
\newcommand{\ev}{\end{verbatim}}
\newcommand{\fr}{\frac}

\newcommand{\hb}{\hbar}

\newcommand{\ka}{\kappa}
\newcommand{\La}{\Lambda}

\newcommand{\na}{\nabla}

\newcommand{\Th}{\Theta}
\newcommand{\th}{\theta}
\newcommand{\vb}{\verb}

\newcommand{\be}{\begin{equation}}
\newcommand{\ee}{\end{equation}} 
\newcommand{\eei}{\end{equation}\indent\indent}
\newcommand{\bc}{\begin{center}}
\newcommand{\ec}{\end{center}}
\newcommand{\ber}{\begin{eqnarray}}
\newcommand{\ear}{\end{eqnarray}}
\newcommand{\ba}{\begin{array}}
\newcommand{\ea}{\end{array}}

\newcommand{\p}{\partial}

\def\case#1/#2{\textstyle\frac{#1}{#2} }
\begin{document}
\title{Exact Nonnull Wavelike Solutions to Gravity with Quadratic Lagrangians.}
\author{Mark D. Roberts, \\\\
Department of Mathematics and Applied Mathematics, \\ 
University of Cape Town,\\
Rondbosch 7701,\\
South Africa\\\\
roberts@gmunu.mth.uct.ac.za} 
\date{\today}
\maketitle
\vspace{0.1truein}
\bc Published:  {\it Int.J.Mod.Phys.} {\bf 9}(1994)167-179.\ec
\bc Eprint: gr-qc/9904007\ec
\bc Comments:  16 pages, no diagrams or tables,  LaTex2e.\ec
\bc 3 KEYWORDS:\ec
\bc Graviational Waves:~~  
    Quadratic Lagrangians:~~  
    Exact Solutions.\ec
\bc 1999 PACS Classification Scheme:\ec
\bc http://publish.aps.org/eprint/gateway/pacslist \ec
\bc 04.30+x,  04.20Jb\ec
\bc 1991 Mathematics Subject Classification:\ec
\bc http://www.ams.org/msc \ec
\bc 83C35,  83C15\ec
\newpage
\artsectnumbering
\begin{abstract}
Solutions to gravity with quadratic Lagrangians are found for the 
simple case where the only nonconstant metric component is the lapse $N$
and the Riemann tensor takes the form $R^{t}_{.itj}=-k_{i}k_{j},~i,j=1,2,3$;
thus these solutions depend on cross terms in the Riemann tensor and therefore
complement the linearized theory where it is the derivatives of the Riemann
tensor that matter.   The relationship of this metric to the null gravitational
radiation metric of Peres is given.   Gravitaional energy Poynting vectors are
construcetd for the solutions and one of these,  based on the Lanczos tensor,
supports the indication in the linearized theory that nonnull gravitational
radiation can occur.
\end{abstract}
\section{Introduction}
\label{sec:intro}
In general relativity it is thought that the assumption of the field equations 
is sufficient to ensure that gravitational radiation is null (i.e. travels at
the speed of light).   It is not necessarily the case that gravitational 
radiation is null \cite{bi:mdr87},  but in general relativity four facts 
support this view.   The {\it first} is that in the linear approximation,  
with the harmonic gauge,  the field equations reduce to $\Box h_{ab}=0$,
which has only null wave solutions.   The weak field linear approximation
may no do justice to the non;inearity of the fielf equations;  for example,
on a de Sitter background,  with cosmological constant $\La$,  
perturbations obey the Fierz-Pauli equation $(\Box-m^2)h_{ab}=0$,  
with mass $m^2=2\La/3$ \footnote{footnote added 1999,  this calculation is
done section 6 gr-qc/9812091}.   The {\it second} fact is that exact solutions
of Einstein's field equations which are good models of gravitational radiation
represent null radiation.   The {\it third} is that the charateristics (or
shock wave soltuions) of the field equations are null;  however,  this 
is not a particularly compelling reason - for example,  the characteristics
of the massive Klein-gordon equation are null \cite{bi:synge}.   
The {\it fourth} is that as a consequence of choosing a metric geometry
(where $\na_cG_{ab}=0$),  there is the equation$\Box g_{ab}=0$,  which 
appears to be massless;  this has led to the construction of nonmetric 
massive theories of gravity obeying the equation $(\Box+m)g_{ab}=0$ 
\cite{bi:mdr86}.

One of the most important variants of general relativity is the theory with
quadratic terms added to the Lagrangian.   Einstein's field equations of 
general relativity can be derived from Hilbert's Lagrangian \cite{bi:hilbert}. 
This Lagrangian was generalized to include terms of higher order in the early 
days of relativity,  by Pauli \cite{bi:pauli},  Bach \cite{bi:bach} and others.
Recently \cite{bi:schmidt} \cite{bi:mdr91},  the relevance of quadratic
Lagrangian theories to the shape of galaxies has been discused.   When 
quadratic Lagrangian theories are linearized \cite{bi:stelle} \cite{bi:BC},
it is found that the field equations reduce to two wave equations with masses
fixed by the values of the coupling constants.   In one case the massis 
positive,  leading to slower-than-light waves,  and in the other it is 
negative,  leading ot faster-than-light waves.   The initial value formulation
of quadratic Lagrangian theory \cite{bi:noakes},  suggests that these masses 
are present in the full nonlinear theory.   Here the existence of exact 
nonnull wavelike solutions to the quadratic Lagrangian theory is investigated.
The method used is to start with a modification of the Peres' wave 
\cite{bi:peres},  where the modification entails changing from Peres' 
dependence on a null coordinate to a nonnull coordinate.   Using of coordinate
transformations,  when the coordinate is timelike the metric is shown to be
equivalent to a metric with the lapse $N$ as the only varying component of
the metric,  and when the metric is spacelike the metric is shown to be static.
The Peres metric itself has been discussed in the context of quadratic 
Lagrangian theories by Buchdahl \cite{bi:buchdahl} and Madsen \cite{bi:madsen};
the metric they use has cylindrical symmetry and depends on a null coordinat,
whereas the solutions here do not have these properties.   The conventions 
used are those of Hawking and Ellis \cite{bi:HE}.
\section{Quadratic Lagrangians}
\label{sec:QL}
The action is taken to be of the form
\be
S=\int g^\fr{1}{2}[{\mathcal L}_m+\ka^{-2}(R-2\La)+bC^2+pR^2]dx^4,
\label{eq:2.1}
\ee
where ${\mathcal L}_m$ is the matter Lagrangian,  the $R$ term is the Hilbert
\cite{bi:hilbert} action,  $\La$ is the cosmological constant,  
the $C^2$ term is the Bach \cite{bi:bach} action with $C^2=C_{abcd}C^{abcd}$ 
being the square of the Weyl tensor,  and the $R^2$ term is the Pauli 
\cite{bi:pauli} action.   Such nomenclature is historically simplistic;  for 
example,  Pauli was considering theories with nonmetric connection.   
Linearization \cite{bi:stelle} \cite{bi:BC} shows that the coupling 
constants can be identified with the masses
\be
\ka^{-2}=\fr{cm^2_{PL}}{16\pi\hb},~~~
m^2_{PL}=\fr{c\hb}{G},~~~
b=\fr{-1}{2\ka^2m^2_{RB}},~~~
p=\fr{+1}{6\ka^2m^2_{WP}}.
\label{eq:2.2}
\ee
For $\La=0$,  there are eight physical degrees of freedom,   two for the 
massless graviton corresponding to the Hilber term,  one massive scalar
corresponding to the Pauli term,  and five massive spin $2$ poltergeists
corresponding to the Bach term.   The sign of these terms can also be 
considered from the point of view of cosmological stability \cite{bi:CT}.

Varying \ref{eq:2.1} gives the field equations
\be
\ka^{-2}(G_{ab}+\La g_{ab})+M_{ab}+X_{ab}=T_{ab},
\label{eq:2.3}
\ee
where
\ber
&&G_{ab}=R_{ab}-\fr{1}{2}g_{ab}R,\label{eq:2.4}\\
&&M_{ab}=-2b\Box R_{ab}
       +2\left(p+\fr{b}{3}\right)R_{;ab}
       -\left(2p-\fr{b}{3}\right)g_{ab}\Box R,\label{eq:2.5}\\
&&X_{ab}=4bR_{acdb}R_{..}^{cd}
       -2\left(p-\fr{2b}{3}\right)RR_{ab}\nonumber\\
&&~~~~~~~~~       +g_{ab}\left[bR_{cd}R_{..}^{cd}
          +\fr{1}{2}\left(p-\fr{2b}{3}\right)R^2\right].
\label{eq:2.6}
\ear
The trace of \ref{eq:2.3} is
\be
\ka^{-2}(4\La-R)-6p\Box R=T,
\label{eq:2.7}
\ee
with $T=T^{~c}_{c.}$,  which leads to the alternative form of the 
field equations
\be
\ka^{-2}(R_{ab}-g_{ab}\La)+X_{ab}+Z_{ab}=S_{ab},
\label{eq:2.8}
\ee
where
\ber
&&Z_{ab}=-2b\Box R_{ab}
         +\left(p+\fr{b}{3}\right)[2R_{;ab}+g_{ab}\Box R],\label{eq:2.9}\\
&&S_{ab}=T_{ab}-\fr{1}{2}g_{ab}T,
\label{eq:2.10}
\ear
and it is this form of the field equations that is used here.   The stress 
tensor is taken to vanish or to be that of a perfect fluid,
\be
S_{ab}=(\mu+\nu)U_aU_b+\fr{1}{2}(\mu-\nu)g_{ab},
\label{eq:2.11}
\ee
where $\mu$ and $\nu$ are respectively the density and the pressure of the 
fluid.   $U_a$ is a unit timelike vector field,
\be
U_aU^a=-1,
\label{eq:2.12}
\ee
with acceleration,  expansion and projection,
\ber
A^a&=&U^a_{~;b}u^b=\dot{U}^a,\nonumber\\
\Th&=&u^a_{~;a},\\
i^{ab}&=&g^{ab}+U^aU^b,\nonumber
\label{eq:2.13}
\ear
respectively.   The first and second conservation equations are
\ber
\mu_aU^a+(\mu+\nu)\Th&=&0,\nonumber\\
(\mu+\nu)A^a+i^{ab}\nu_b&=&0,
\label{eq:2.14}
\ear
respectively.
\section{Poynting Vectors}
\label{sec:PV}
In a nonvacuum space-time the speed of a gravitational wave can be measured
relative to the matter present.   For example,  in a cosmological model it 
can be compared to the speed of the comoving fluid velocity vector;  on a 
galactic scale the speed of a gravitational wave emitted from the center of 
the galaxy could be estimated be comparing it to the speed of co-occurring 
electromagnetic radiation,  or possibly by direct methods.   The metrics that 
will be here have the property \ref{eq:6.1},  where the first conservation
implies that the density of a fluid is at rest in this coordinate system,  so
that there is a possiblity of measuring the speed of a gravitational wave 
with respect to this fluid.   So far the field equations coupled to a perfect 
fluid have proved intractable,  and only solutions with zero stress have been 
found;  thus for a timelike wave it is possible to move to a coordinate 
system where the wave appears to be at rest,  and as there is no matter 
present this criterion cannot be used to measure wave speed.   Another 
criterion is needed and the one suggested here is to construct Poynting 
vectors representing the speed of energy transfer:  whether the Poynting
vector is timelike,  null  or spacelike will be taken to support the belief
that the gravitational wave is timelike,  null or spacelike.   If the 
principle of equivalence holds,  then the energy of the gravitaional field
appears to be nonexistent to an observer in free fall,  and consequently a
Poynting vector for gravitaional energy gives at best only an indication of 
the real energetics involved.   To sum up:  whether a gravitational enery
Poynting vector is timelike,  null or spacelike can be regarded as measuring 
whether the gravitational wave is timelike,  null or spacelike;  alternatively
it can be considered as a measure of the appropriatenees,  or otherwise, of
the Poynting vector.

There are tensors constructed out of products and derivatives of the Riemann 
tensor,  which can represent the square of gravitational energy \cite{bi:bel}
There are a large number of possible combinations \cite{bi:collinson}.
An often-used combination is produced by first Matt\'e-decomposing the Weyl 
tensor
\ber
C_{abcd}=R_{abcd}&+&\fr{1}{6}R(g_{ac}g_{db}-g_{ad}g_{cb})\nonumber\\
        &+&\fr{1}{2}(g_{ad}R_{cb}-g_{ac}R_{db}+g_{bc}R_{da}-g_{db}R_{ca}),
\label{eq:3.1}
\ear
into its electric and magnetic parts,
\be
E_{ab}\equiv C_{ambn}U^mU^n,~~~
B_{ab}\equiv *C_{ambn}U^mU^n,
\label{eq:3.2}
\ee
where $U_a$ is a timelike vector field and 
$*C_{abcd}=\fr{1}{2}\sqrt{-g}\ep_{abcd}C^{mn}_{..cd}$,  and then defining
\be
T_{abcd}\equiv C_{ambn}C^{~m~n}_{c.d.}+*C_{ambn}*C^{~m~n}_{c.d.},
\label{eq:3.3}
\ee
as an energy squared tensor,  to give the Poynting vector
\ber
P^a&\equiv&T^a_{.bcd}U^bU^cU^d\nonumber\\
   &=&C^a_{.mbn}E^{mn}_{..}U^b+*C^a_{.mbn}B^{mn}_{..}U^b.
\label{eq:3.4}
\ear

Energy tensors of the correct dimension can be constructed using the Lanczos
tensor \cite{bi:mdr88}.   Again there are a large number of possibilities,
with the added complication of the choice of gauges possible for the Lanczos 
tensor.   Here it is assumed that both the Lanczos algebraic
\be
3\ch_a\equiv H^{~b}_{a.b}=0,
\label{eq:3.5}
\ee
gauge and differential
\be
D_{ab}\equiv H^{~~c}_{ab.;c},
\label{eq:3.6}
\ee
gauge ave been applied.  Subject to these gauges the Weyl tensor is recovered
as a linear diferential equation in the Lanczos tensor
\ber
C_{abcd}&=&H_{abc;d}-H_{abd;c}+H_{cda;b}-H_{cdb;a}\nonumber\\
        &-&g_{ac}H^{~e}_{b.d;e}+g_{ad}H^{.e}_{b.dc;e}
         -g_{bd}H^{~e}_{a.c;e}+g_{bc}H^{~e}_{a.d;e}.
\label{eq:3.7}
\ear
The Lanczos tensor has the symmetries
\ber
&&H_{abc}=-H_{bac},\nonumber\\
&&H_{abc}=H_{cba}-H_{cab}.
\label{eq:3.8}
\ear
Two of the more successfu; energy tensors are
\ber
L_{ab}&=&H^{~cd}_{a..}H_{bcd}+*H^{~cd}_{a..}*H_{bcd}\nonumber\\
      &=&2H^{~cd}_{a..}H_{bcd}-\fr{1}{2}g_{ab}H_{cde}H^{cde}_{...},\\
\label{eq:3.9}
M_{ab}&=&2H^{~c}_{a.b;c}.
\label{eq:3.10}
\ear
Their Poynting vectors can be defined as
\ber
O^a_.&\equiv&L^{ab}_{..}U_b,\nonumber\\
Q^a_.&\equiv&M^{ab}_{..}U_b,
\label{eq:3.11}
\ear
where $U_a$ is a timelike vector.
\section{The Metric}
\label{sec:metric}
The line element is taken to be of the form
\be
ds^2=-dt^2+dx^2+dy^2+dz^2+h(u,x,y)du^2,
\label{eq:4.1}
\ee
where
\be
u\equiv lt+z.
\label{eq:4.2}
\ee
For $l=1$,  $u$ is null and this is Peres' metric;
for $l^2<1$ or $l^2>1$,  $u$ is spacelike or timelike respectively.
For $l^2<1$ define
\be
\sqrt{1-l^2}\cdot z'\equiv lt+z\equiv u,~~~
\sqrt{1-l^2}\cdot t'\equiv t+lz.
\label{eq:4.3}
\ee
Then the metric becomes
\be
ds^2=-dt^2+dx^2+dy^2+[1+(1-l^2)h(z',x,y)]dz'^2,
\label{eq:4.4}
\ee
which is manifestly static.   For $l^2>1$ define
\be
\sqrt{l^2-1}\cdot t''\equiv lt+z\equiv u,~~~
\sqrt{l^2-1}\cdot z''\equiv t+lz.
\label{eq:4.5}
\ee
Then the metric becomes
\be
ds^2=-[1+(1-l^2)h(\sqrt{l^2-1}\cdot t'',x,y)]dt''^2+dx^2+dy^2+dz''^2.
\label{eq:4.6}
\ee
This metric has the lapse $N$ as the only varying component of the metric and
may be written in the form
\be
ds^2=-N^2dt^2+dx^2+dy^2+dz^2.
\label{eq:4.7}
\ee
When the lapse is of the variables seperable form
\be
N=H(t)\al(x,y),
\label{eq:4.8}
\ee
the time dependence can be absorbed into the metric,
showing that the metric is static.
\section{The Field Equations}
\label{sec:fe}
The line element \ref{eq:4.7} gives the Christoffel symbols
\be
\vb+{+^t_{tt}\vb+}+=\fr{\dot{N}}{N},~~~
\vb+{+^t_{ti}\vb+}+=\fr{N_i}{N},~~~
\vb+{+^i_{tt}\vb+}+=NN^i_.,
\label{eq:5.1}
\ee
where $i,j=x,y,z$ and $\dot{N}=\p-tN$.   The Riemann tensor,  Ricci tensor
and Ricci scalar have nonvanishing components:
\ber
R^t_{.itj}&=&-\fr{N_{ij}}{N},~~~R^i_{.tjt}=NN^i_{.j},\label{eq:5.2}\\
R_{tt}&=&NN^{~i}_{i.},~~~R_{ij}=-\fr{N_{ij}}{N},\label{eq:5.3}\\
R&=&-2\fr{N^{~i}_{i.}}{N}.\label{eq:5.4}
\ear
The first covariant derivatives of the Riemann tensor are
\ber
R_{itjt;t}&=&N^2\left(\fr{N_{ij}}{N}\right)^\circ,~~~
R_{itjt;k}=N^2\left(\fr{N_{ij}}{N}\right)_k,\label{eq:5.5}\\
R_{jtik;t}&=&N_iN_{jk}-N_kN_{ji}.\nonumber
\ear
The first covariant derivatives of the Ricci tensor can be found 
by contraction;  note in particular
\be
R_{it;t}=N^j_.N_{ij}-N_iN^{~j}_{j.}.
\label{eq:5.6}
\ee
The second covariant derivatives of the Riemann tensor are
\ber
R_{itjt;tt}&=&N^3
\left\{\left[\fr{1}{N}\left(\fr{N_{ij}}{N}\right)^\circ\right]^\circ
          -\left[N\left(\fr{N_{ij}}{N}\right)_k\right]^k\right\}\nonumber\\
&&+N^4\left(\fr{N_{ij}}{N}\right)^{~k}_{k.}
-NN^k(R_{jtik;t}+R_{itjk;t}),\nonumber\\
R_{jtik;tt}&=&N^2\left[\fr{1}{N^2}(n_iN_{jk}-N_kN_{ji}\right]^\circ\nonumber\\
         &&-N\left[N_i\left(\fr{N_{jk}}{N}\right)^\circ
                 +N_k\left(\fr{N_{jk}}{N}\right)^\circ\right],\\
R_{ijkl;tt}&=&\fr{2}{N}(N-iN-kN_{jl}-N_iN_lN_{jk}
                     -N_jN_kN_{ij}+N_lN_jN_{ik}),\nonumber\\
R_{itjt;kl}&=&N^2\left(\fr{N_{ij}}{N}\right)_{kl}.\nonumber
\label{eq:5.7}
\ear
The tensors \ref{eq:2.9} and \ref{eq:2.6} are
\ber
Z_{tt}=&2&\left(p-\fr{2b}{3}\right)N
          \left[Ng^{ab}\left(\fr{N^{~i}_{i.}}{N}\right)_a\right]_b
          -\fr{4bN^i_.}{N}R_{it;t}\nonumber\\
       &+&4\left(p+\fr{b}{3}\right)NN^i_.
            \left(\fr{N^{~j}_{j.}}{N}\right)_i,\nonumber\\
Z_{it}=&-&2b\left[-N\left(\fr{R_{it;t}}{N^3}\right)^\circ+\fr{1}{N^2}
            (N_i\dot{N}^{~j}_{j.}-N^j_.\dot{N}_{ij})\right]\nonumber\\
       &-&4\left(p+\fr{b}{3}\right)N
            \left[\fr{1}{N}\left(\fr{N^{~j}_{j.}}{N}\right)^\circ\right]_i\\
Z_{ij}=&-&2b\left\{\fr{1}{N}\left[Ng^{ab}
                             \left(\fr{N_{ij}}{N}\right)_a\right]_b
           -\fr{1}{N^3}(N_iR_{jt;t}+N_jR_{it;t})\right\}\nonumber\\
       &+&\left(p+\fr{b}{3}\right)\left\{
                 -4\left(\fr{N^{~k}_{k.}}{N}\right)_{ij}
               -\et_{ij}\fr{2}{N}\left[Ng^{ab}
                   \left(\fr{N^{~k}_{k.}}{N}\right)_a\right]_b\right\},
\label{eq:5.8}
\ear
and
\ber
X_{tt}&=&3bN_{ij}N^{ij}_{..}
           +\left(2p-\fr{7b}{3}\right)N^{~i}_{i.}N^{~j}_{j.},\\
X_{it}&=&0,\nonumber\\
X_{ij}&=&-4p\left(p+\fr{b}{3}\right)\fr{N_{ij}N^{~k}_{k.}}{N^2}
         +\fr{\et_{ij}}{N^2}\left(bN_{kl}N^{kl}_{..}
             +\left(2p-\fr{b}{3}\right)n^{~l}{l.}N^{~k}_{k.}\right),\nonumber
\label{eq:5.9}
\ear
respectively.
\section{Solutions to the Field Equations}
\label{sec:sol}
From vacuum general relativity,  comparing the Ricci tensor \ref{eq:4.3}
and the Riemann tensor \ref{eq:4.2} we see that the only solution 
is flat space.   For a perfect fluid the first and second conservation 
equations \ref{eq:2.14} become
\ber
\mu^0&=&0,\nonumber\\
(\mu+\nu)\fr{N_i}{N}&+&\nu_i=0,
\label{eq:6.1}
\ear
respectively.   The stress is given by
\ber
S^{~t}{t.}&=&-\fr{1}{2}(3\mu+\nu),\nonumber\\
S_{ij}&=&\fr{1}{2}\et_{ij}(\mu-\nu).
\label{eq:6.2}
\ear
For the perfect fluid in general relativity,  the $S_{xy},  S_{xz},  S_{zy}$ 
equations show that $N$ is a function of one spatial coordinate,  say $x$,
and then comparing $S_{xx}$ and $S_{yy}$ shows that this vanishes;  again
the only solution is flat space.

In general the firld equations constructed in Sec \ref{sec:fe} are too 
complex to solve,   especially because of the occurrence of $R_{it;t}$.
Taking
\be
x=r~sin \th,~~~
y=r~cos \th,
\label{eq:6.3}
\ee
gives
\ber
&&R_{rt;t}=\fr{1}{r^2}\left(N_\th N_{\th,r}
                            -\fr{N^2_\th}{rN}-N_rN_{\th,\th}
                            +2rN^2_r\right)
              +N_zN_{rz}-N_rN_{zz},\nonumber\\
&&R_{\th t;t}=N_rN_{\th,r}-\fr{N_rN_\th}{rN}-N_\th N_{,rr}
                          +N_zN_{\th z}-N_\th N_{zz},\\
&&R_{zt;t}=N_rN_{zr}-N_zN_{rr}
           +\fr{1}{r^2}[N_\th N_{z\th}-N_z(N_{\th\th}-2rN_r)].\nonumber
\label{eq:6.4}
\ear
Requiring $N=N(t,r)$ still leaves $R_{rt;t}$ nonvanishing.
$R_{it;t}$ can be made to vanish by assuming that
\be
N=N(t,k-ix^i).
\label{eq:6.5}
\ee
Then if one lets $k^2=k_ik^i$ the field equations become
\ber
S^{~t}_{.t}&=&-\fr{\La}{\ka^2}-\fr{k^2N''}{\ka^2N}
            -\fr{2}{N}\left(p-\fr{2b}{3}\right)
              \left[Ng^{ab}\left(\fr{N''}{N}\right)_a\right]_b\nonumber\\
           && +2k^2\left(p+\fr{b}{3}\right)
 \left\{\fr{2}{N}\left[\fr{1}{N}\left(\fr{N''}{N}\right)^\circ\right]^\circ
-\fr{k^2}{N^2}\left[N''^2
                   +2NN'\left(\fr{N''}{N}\right)'\right]\right\},\nonumber\\
S^{~i}_{t.}&=&-4k^2K^i_.\left(p+\fr{b}{3}\right)
               N\left[\fr{1}{N}\left(\fr{N''}{N}\right)^\circ\right]',\\
S^{~j}_{i.}&=&k_ik^j_.\left\{-\fr{N''}{\ka^2N}
               +\fr{2b}{N}\left[Ng^{ab}\left(\fr{N''}{N}\right)_a\right]_b
              -4k^2\left(p+\fr{b}{3}\right)\left[\left(\fr{N''}{N}\right)''
                                  +\fr{N''^2}{N^2}\right]\right\}\nonumber\\
           &&+2\left(p+\fr{b}{3}\right)\de^{~j}_{i.}k^2\left\{-\fr{1}{N}
 \left[Ng^{ab}\left(\fr{N''}{N}\right)_a\right]_b+\fr{k^2N''^2}{N^2}\right\}
                -\fr{\La}{\ka^2}\de^{~j}_{i.}.\nonumber
\label{eq:6.6}
\ear

First consider the case
\be
p+\fr{b}{3}=\fr{1}{6\ka^2}\left(\fr{1}{m^2_{WP}}-\fr{1}{m^2_{RB}}\right)=0,
\label{eq:6.7}
\ee
and $\La=0$;  then the field equations become
\ber
S_{ab}&=&-2b\left(\Box-\fr{1}{2b\ka^2}\right)R_{ab}\nonumber\\
      &=&\fr{1}{\ka^2m^2_{RB}}(\Box+m^2_{RB})R_{ab}.
\label{eq:6.8}
\ear
For $S_{ab}=0$,  by the use of \ref{eq:5.3} and \ref{eq:5.7} this equation is
\be 
0=\left[\fr{1}{N}\left(\fr{N_{ij}}{N}\right)^\circ\right]^\circ
  -\left[N\left(\fr{N_{ij}}{N}\right)_k\right]^k
  -m^2_{RB}N_{ij},
\label{eq:6.9}
\ee
where $\dot{N}=\p_tN$.   This has the static solution
\be
N=A(t)\exp\left(a_0+a_1z-\fr{1}{4}m^2_{RB}z^2\right),
\label{eq:6.10}
\ee
where $a_0$ and $a_1$ are constants.   This solution can be rotated around
the spatial axis,  and also transferred into the form \ref{eq:4.1}.

A nonstatic solution can be found by assuming that the lapse $N$ is a 
function of $w=k_ax^a$,  where $k_a$ is a constant.   Symbolically this is
\be
N=N(-k_0t+k_ix^i)=N(k_ax^a)=N(w).
\label{eq:6.11}
\ee
The Riemann tensor takes the simple form
\be
R^t_{.itj}=-k_ik_j\fr{N''}{N}.
\label{eq:6.12}
\ee
If one defines
\be
t\equiv u=\fr{lt'+z'}{\sqrt{l^2-1}},~~~
z\equiv\fr{t'+lz'}{\sqrt{l^2-1}},
\label{eq:6.13}
\ee
the metric can be put in the form \ref{eq:3.1} with
\be
h=1-N^2\left[u\left(-k_0+\fr{k_z}{l}\right)+
             k_xx+k_yy+k_z\fr{\sqrt{l^2-1}}{l}z'\right].
\label{eq:6.14}
\ee
If one assumes that $p=-b/3$ as in \ref{eq:6.7},  the field equations are
\ber
S^{~t}_{t.}=S^{~i}_{i.},~~~~~~~~~~~~~~~~~~~~~~~~~~~~~~~~~~~~~~\nonumber\\
\ka^2m_2^2S_{ij}=k_ik_j\left\{-m^2_2\fr{N''}{N}
      -\fr{k_0^2}{N}\left[\fr{1}{N}\left(\fr{N''}{N}\right)'\right]'
      +\fr{k^2}{N}\left[N\left(\fr{N''}{N}\right)'\right]'\right\}.
\label{eq:6.15}
\ear
When $S_{ab}=0$ these equations reduce to the single equation
\be
-\fr{k^2_0}{N}\left[\fr{1}{N}\left(\fr{N''}{N}\right)'\right]'
+\fr{k^2}{N}\left[N\left(\fr{N''}{N}\right)'\right]'-m^2_2\fr{N''}{N}=0.
\label{eq:6.16}
\ee
Integrating once,  with the constant of integration $a_1$,
and multiplying by $N$ gives
\be
(k^2N^2-k_0^2)\left(\fr{N''}{N}\right)'-m_2^2NN'-a_1N=0.
\label{eq:6.17}
\ee
Setting the constant of integration $a_1=0$ and integrating again with
the constant of integration $a_2$ gives
\be
\fr{N''}{N}=a_2+\fr{1}{2}m_2^2k^{-2}\ln(k^2N^2-k_0^2).
\label{eq:6.18}
\ee
This differential equation remains intractable when $m_2\ne0$,
but when $m_2=0$ it has the simple solution
\be
N=C_+\exp(k_ax^a)+C_-\exp(-k_ax^a),
\label{eq:6.19}
\ee
where a factor of $\sqrt{a_2}$ has been absorbed into $k_a$,
and thus $\sqrt{a_2}k_a\rightarrow k_a$;
because $a_2$ is a constant of integration $k_a$ may now be complex.
For \ref{eq:6.19} the Riemann tensor takes the simple form
\be
R^t_{.itj}=-k_ik_j.
\label{eq:6.20}
\ee

For $p+\fr{b}{3}\ne0$,  inspection of the $S^{~i}_{t.}$ term \ref{eq:6.6} 
and the fluid conservation equations \ref{eq:6.1} suggests that it should be 
possible to integrate the field equations,  
but so far this has turned out not to be the case.
It is known \cite{bi:mdr89} that a sperically symmetric metric 
is always a solution of the general relativity field equations 
if a sufficiently large number of other fields and fluids are present.
Adding more terms to the stress in \ref{eq:6.6} 
has not yet given invertible equations.
To find $p+\fr{b}{3}\ne0$ solutionsthe expresion \ref{eq:6.20},
i.e.  $R^t_{.itj}=-k_ik_j=-N_{ij}/N$,  is substituted back
into the field equations \ref{eq:6.6}.   
This assumption amounts to disregarding higher derivatives of the Rieman 
tensor and investigating only the effect of cross terms;
therefore it complements previous linear analysis.
Substituting \ref{eq:6.6} gives the field equations
\ber
S^{~t}_{t.}&=&-\fr{k^2}{\ka^2}
              -\fr{\La}{\ka^2}
              -2k^4\left(p+\fr{b}{3}\right),\nonumber\\
S^{~i}_{t.}&=&0,\\
S^{~i}_{i.}&=&-\fr{k^2}{\ka^2}
              -\fr{3\La}{\ka^2}
              +2k^4\left(p+\fr{b}{3}\right),\nonumber\\
S^{~j}_{i.}(i\ne j)&=&k_ik^j\left[-\fr{1}{\ka^2}
                     -4k^2\left(p+\fr{b}{3}\right)\right].\nonumber
\label{eq:6.21}
\ear
Taking $S_{ab}=0$ gives the solution
\ber
k^2&=&-2\La=\fr{-1}{4\ka^2\left(p+\fr{b}{3}\right)}\nonumber\\
   &=&\fr{-3}{2(m^{-2}_{WP}-m^{-2}_{RB})}.
\label{eq:6.22}
\ear
\section{Poynting Vectors for the Metric}
\label{sec:Poy}
From Eqs. \ref{eq:3.1} and \ref{eq:4.2} the Weyl tensor is found to be
\ber
C^t_{.itj}&=&-\fr{N_{ij}}{2N}+\fr{N^{~k}_{k.}}{6N}\et_{ij},\nonumber\\
C^i_{jkl}&=&\fr{1}{2N}(-\de^i_lN_{jk}+\de^i_kN_{jl}
                     -\et_{jk}N^i_l+\et_{jl}N^i_k)\\
           &&-\fr{N^{~m}_{m.}}{3N}(\de^i_k\et_{jl}-\de^i_l\et_{jk}).\nonumber
\label{eq:7.1}
\ear
The electric and magnetic parts become
\ber
E_{ij}&=&\fr{N_{ij}}{2N}-\fr{N^{~k}_{k.}}{6N}\et_{ij},\nonumber\\
B_{ab}&=&0.
\label{eq:7.2}
\ear
The Poynting vector \ref{eq:3.4} is
\be
P^a_.=\fr{\de^a_t}{12N^3}(3N_{ij}^{~~2}-N_{ii}^{~~2}),
\label{eq:7.3}
\ee
and thus
\be
P_aP^a=\fr{-1}{12^2N^4}(3N_{ij}^{~~2}-n_{ii}^{~~2})^2,
\label{eq:7.4}
\ee
is always timelike.   This Poynting vector appears to give no indication of 
energy transfer as $P^t_.$ is the only nonvanishing component.

The Lanczos tensor can be calulated by direct methods similar to those used
in Ref.\cite{bi:mdr95}.   In the Lanczos algebraic and differential gauges 
it is found to be
\be
H^t_{.it}=-\fr{N_i}{3N},~~~
H_{ijk}=\fr{1}{6N}(N_j\et_{ik}-N_i\et_{jk}).
\label{eq:7.5}
\ee
The energy tensor \ref{eq:3.9} is
\be
L_{tt}=-\fr{N_i^2}{18},~~~
L_{it}=0,~~~
L_{ij}=\fr{1}{18N^2}(5N_iN_j-2\et_{ij}N^2_k),
\label{eq:7.6}
\ee
and the associated Poynting vector is
\be
O^a_.=-\fr{N^2_i\de^a_t}{18N^2};
\label{eq:7.7}
\ee
thus
\be 
O_aO^a=-\left(\fr{N_i^2}{18N^2}\right)^2,
\label{eq:7.8}
\ee
and again this is always timelike.

The energy tensor \ref{eq:3.10} is
\ber
M^t_{.t}&=&-\fr{2}{3}\left(\fr{N^i_.}{N}\right)_i,~~~
M^i_{.t}=-\fr{2}{3}\left(\fr{N^i_.}{N}\right)^\circ,\\
M_{ij}&=&\fr{2N_iN_j}{3N^2}-\fr{N_{ij}}{3N}+\et_{ij}\fr{N^k_{.k}}{3N},\nonumber
\label{eq:7.9}
\ear
and the associated Poynting vector is
\be
Q^a_.=\fr{2}{3N}\left[
        \left(\fr{N^i_.}{N}\right)_i,\left(\fr{N^i_.}{N}\right)^\circ\right],
\label{eq:7.10}
\ee
giving
\be
Q_aQ^a=-\fr{4}{9}g^{ab}\left(\fr{N^i_.}{N}\right)_a\left(\fr{N_i}{N}\right)_b.
\label{eq:7.11}
\ee
For \ref{eq:6.20} with either $C_+=0$ or $C_-=0$,  $Q^a=0$;
however,  for $C=C_+=C_-$,
\be
Q_aQ^a=-\fr{4}{9}k^2\left(k^2-\fr{k_0^2}{N^2}\right)Sech^4(k_ax^a).
\label{eq:7.12}
\ee
For the solution \ref{eq:6.22},  \ref{eq:7.12} reduces to
\be
Q_aQ^a=\fr{-1}{(m^{-2}_{WP}-m^{-2}{_RB})}
        \left(\fr{1}{m^{-2}_{WP}-m^{-2}_{RB}}+\fr{2k_0^2}{3N^2}\right)
        \cdot Sech(k_ax^a).
\label{eq:7.13}
\ee
When $m_{RB}=0$,  $Q^a$ is always timelike;  when $M_{RB}\ne 0$ a variety of
behaviour is possible,  depending on the values of the constants.
\section{Acknowledgements}
I would like to thank the Leverhulme Trust for financial support.

\end{document}